# 基于局部学习机和细菌群体趋药性算法的电力系统暂态稳定评估

顾雪平 李 扬

（华北电力大学电气与电子工程学院 保定 071003）

**摘要** 为了提高电力系统暂态稳定评估的分类正确率，提出一种基于局部学习机（LLM）和改进的细菌群体趋势药性（BCC）算法的暂态稳定评估方法。该方法采用 LLM 构建暂态稳定评估模型，考虑相量测量单元可以提供的故障后实测信息，构造了一组系统特征作为 LLM 模型的输入量，稳定结果作为输出量，对稳定结果和系统特征间的映射关系进行训练，并通过综合混沌搜索策略的改进 BCC 算法优化 LLM 模型的参数。最后，以新英格兰 10 机 39 节点系统为例验证了所提方法的有效性。

**关键词**：暂态稳定评估 局部学习机 细菌群体趋药性 混沌搜索 相量测量单元

**中图分类号**：TM74

## Transient Stability Assessment of Power Systems Based on Local Learning Machine and Bacterial Colony Chemotaxis Algorithm

*Gu Xueping   Li Yang*

（North China Electric Power University Baoding 071003 China）

**Abstract** In order to improve the classification accuracy of transient stability assessment of power systems, a novel method based on local learning machine and an improved bacterial colony chemotaxis (BCC) algorithm is proposed, where local learning machine(LLM) is used to build a TSA model. Considering the possible real-time information provided by PMU, a group of system-level classification features extracted from the power system operation parameters are employed as inputs, and the stability result is used as output of the LLM model. The relation ship between input and output is trained and the ideal model is obtained by applying the improved BCC combined with chaotic search strategy to determine the optimal parameters of LLM automatically. The effectiveness of the proposed method is shown by the simulation results on the New England 10-unit-39-bus power system.

**Keywords**：transient stability assessment(TSA), local learning machine, bacterial colony chemotaxis, chaotic search, phasor measurement unit

## 1 引言

电力系统暂态稳定评估（Transient Stability Assessment, TSA）一直是关系到电力系统安全稳定运行的重要问题[1]。近年来，基于人工神经网络（Artificial Neural Network，ANN）、决策树（Decision Tree, DT）、支持向量机（Support Vector Machine, SVM）等机器学习技术的 TSA 受到各国学者的广泛关注，取得了较大的进展[2-8]。该类方法无需建立系统的数学模型，其主要任务是建立系统变量和系统稳定结果间的关系映射,具有学习能力强、评估速度快、能提供潜在有用信息等优势[1]，在电网在线安全稳定分析领域有着较好的应用前景。ANN 是最早应用于 TSA 的人工智能方法之一[2]，但其存在过拟合现象，且模型参数选择困难、训练结

果不够稳定。决策树由比利时学者 L. Wehenkel 首先引入 TSA 中，由于其评估结果为显式的规则集，因而引起了许多研究人员的兴趣[3]，但其评估结果对样本构成非常敏感，而且决策知识的外延能力和鲁棒性较差[1]。SVM 是一种基于统计学习理论的、较新的机器学习方法，它有效克服了 ANN 的过拟合问题[4]，较好地控制了评估模型的泛化误差，因而成为目前该领域中研究最为活跃的方法之一[4,5]，但其存在参数选取困难[7]等不足。最近，文献[6]提出了一种基于概率神经网络（Probabilistic Neural Network，PNN）的 TSA 方法；文献[7]提出了一种基于极限学习机（Extreme Learning Machine, ELM）的 TSA 模型；文献[8]提出了一种基于马尔可夫链蒙特卡罗方法的电力系统暂态稳定概率评估方法。

LLM 是由文献[9]提出的一种用于模式识别特征选择和分类的算法，已被成功应用于生物信息学和生物医学工程等领域[9]，其核心思想是通过局部学习把任意非线性的问题分解为一组局部线性的子问题，然后在大间隔框架内整体学习特征的相关性。选取适当的参数对 LLM 模型的学习性能和泛化能力至关重要，但目前其参数的选取多依赖于经验。BCC 算法具有全局性、快速性和高精度等特点，已被成功用于解决电气设备缺陷参数定量识别、最小二乘支持向量机模型参数的优化选取等问题[10]。

本文提出一种基于局部学习机和改进细菌群体趋势药性算法（Improved BCC, IBCC）的暂态稳定评估方法，并以新英格兰 10 机 39 节点系统为例验证所提方法的有效性。

## 2 LLM 算法简介

LLM 算法基于大间隔理论，其基本思路是对每个特征进行加权，得到一个以非负向量 $w$ 为参数的加权特征空间，使得在该特征空间中基于分类间隔的误差函数最小[9]。

### 2.1 训练阶段

给定训练集 $D=\{(x_n,y_n)\}_{n=1}^{N}\subset R^J \times \{\pm 1\}$，其中 $x_n$ 是包含 $J$ 个特征的第 $n$ 个数据样本的特征向量，$y_n$ 是相应的类标。

#### 2.1.1 分类间隔的定义

给定一个距离函数 $d(\cdot)$（文中采用曼哈顿距离），每个样本 $x_n$ 总可以找到一个同类样本中距离最近的样本 $NH(x_n)$ 和一个异类样本中距离最近的样本 $NM(x_n)$。$x_n$ 的分类间隔与向量 $w$ 有关，其定义如下：

$$\rho_n(w)=d(x_n,NM(x_n)|w)-d(x_n,NH(x_n)|w)=w^T z_n \quad (1)$$

其中，$z_n=|x_n-NM(x_n)|-|x_n-NH(x_n)|$。

因为训练前给定样本在加权特征空间中距离最近的同类、异类样本都是未知的，因此，LLM 算法用一个概率模型计算在定义局部信息中的不确定性。根据期望最大化算法，分类间隔的估计值可通过计算 $\rho_n(w)$ 的期望获得，其定义如下：

$$\bar{\rho}_n(w)=w^T\left[\sum_{i\in M_n}P(x_i=NM(x_n)|w)|x_n-x_i|-\sum_{i\in H_n}P(x_i=NH(x_n)|w)|x_n-x_i|\right]=w^T \bar{z}_n \quad (2)$$

其中

$$M_n=\{i:1\leq i\leq N, y_i\neq y_n\}$$
$$H_n=\{i:1\leq i\leq N, y_i=y_n, i\neq N\}$$

$P(x_i=NM(x_n)|w)$ 和 $P(x_i=NH(x_n)|w)$ 分别表示样本 $x_i$ 是距离样本 $x_n$ 最近的异类样本和同类样本的概率。这些概率通过标准的核密度估计得到。

$$P(x_i=NM(x_n)|w)=\frac{k(\|x_n-x_i\|_w)}{\sum_{j\in M_n}k(\|x_n-x_j\|_w)} \quad \forall i\in M_n \quad (3)$$

$$P(x_i=NH(x_n)|w)=\frac{k(\|x_n-x_i\|_w)}{\sum_{j\in H_n}k(\|x_n-x_j\|_w)} \quad \forall i\in H_n \quad (4)$$

式中，$k(\cdot)$ 是核函数。本文中核函数选用指数函数

$$k(d)=\exp\left(-\frac{d}{\sigma}\right) \quad (5)$$

式中，$\sigma$ 为核宽度参数，用于控制数据被局部分解的细化程度。

#### 2.1.2 特征权重的确定

基于大间隔框架，最优化问题可以表述为

$$\min_{w}\sum_{n=1}^{N}\log(1+\exp(-w^T\bar{z}_n))+\lambda\|w\|_1 \quad w\geq 0 \quad (6)$$

式中，$\lambda$ 为正则化参数，用于控制惩罚力度及特征权重的稀疏化程度。

对于固定的 $\bar{z}_n$，式（6）是一个带约束的凸优化问题，将其改写为

$$\min_{v}\sum_{n=1}^{N}\log\left(1+\exp\left(-\sum_{j}v_j^2\bar{z}_n(j)\right)\right)+\lambda\|v\|_2^2 \quad (7)$$

从而转化为一个无约束优化问题。易知，$w_j = v_j^2$，$1 \leq j \leq J$。然后通过梯度下降法容易求得 $v$，更新规则为

$$v \leftarrow v - \eta \left[ \lambda \mathbf{1} - \sum_{n=1}^{N} \frac{\exp\left(-\sum_j v_j^2 \bar{z}_n(j)\right)}{1 + \exp\left(-\sum_j v_j^2 \bar{z}_n(j)\right)} \bar{z}_n \right] \otimes v \quad (8)$$

式中，$\otimes$ 是 Hadamard 算子；$\eta$ 是由标准线性搜索确定的学习速率。

### 2.2　测试阶段

确定新模式 $x^*$ 的类别时，首先把 $x^*$ 和训练阶段得到的权重 $w^*$ 代入式（2）得到分类间隔 $\bar{\rho}_n^*(w^*)$，然后根据下面的决策函数确定 $x^*$ 所属类别，即

$$f(x) = \text{sgn}\left(\frac{1}{1 + \exp(\bar{\rho}_n^*(w^*))} - 0.5\right) \quad (9)$$

若 $f(x) > 0$，则 $x^*$ 属于第一类；反之，则属于第二类。

由上述训练过程可知，正则化参数 $\lambda$ 和核宽度 $\sigma$ 对 LLM 模型的学习性能和泛化能力有重要影响。在本文所提方法中，这些参数通过 IBCC 算法自动优化选取，以优化 LLM 模型的性能。

## 3　基于 IBCC-LLM 的暂态稳定评估

### 3.1　BCC 算法的改进

BCC 算法是一种个体细菌间的信息交互细菌趋药性算法，它根据单个细菌对化学诱剂的应激反应和细菌群落间的位置信息交互进行优化[10]。

"无免费午餐"定理[11]说明，没有一种算法对任何问题都是最优，即全局寻优能力和收敛速度都最强。BCC 算法中信息交互机制的引入在加快算法收敛速度的同时，也不可避免地带来了早熟收敛和局部最优问题。由于混沌运动具有遍历性、随机性和对初始条件敏感性等特点，使混沌处理方法被广泛用于处理此类优化问题[12]。

本文提出综合混沌搜索策略的改进 BCC 算法，对 BCC 算法从两方面进行改进：一方面对感知范围进行自适应调整，以改善算法性能；另一方面结合混沌思想在 BCC 算法发生早熟收敛时进行混沌搜索，以增强算法跳出局部最优解的能力。

#### 3.1.1　感知范围的自适应调整

在 BCC 算法中，感知范围的合理设置对算法性能有较大影响。感知范围越大，细菌聚集得越快，越容易陷入局部最优；感知范围越小，细菌受环境影响越小，越要靠单个细菌寻优，收敛速度越慢。本文采用感知范围自适应调整的方式，使感知范围随着种群聚集度的提高而减小，从而保持种群多样性，改善算法性能。

在寻优过程中，群体适应度方差能反映种群的聚集程度，适合用来对控制参数进行动态调整，适应度方差 $\sigma^2$ 的计算公式为[12]

$$\sigma^2 = \sum_{i=1}^{N_p} \left(\frac{f_i - f_{\text{avg}}}{f_{\text{best}}}\right)^2 \quad (10)$$

式中，$f_i$ 是第 $i$ 个细菌的适应度值；$f_{\text{avg}}$ 是当前群体的平均适应度；$f_{\text{best}}$ 为群体最佳适应度；$N_p$ 为种群规模。

考虑种群分布情况后，BCC 算法中感知范围 $S$ 的自适应调整策略为

$$S_k = S_{\min} + (S_{\max} - S_{\min}) \times \frac{\sigma_k^2}{N_p} \quad (11)$$

式中，$S_k$ 为第 $k$ 次的感知范围；$S_{\min}$ 和 $S_{\max}$ 分别为感知范围的最小值和最大值。

#### 3.1.2　混沌搜索策略

针对 BCC 的早熟收敛问题，本文通过监视群体适应度方差的变化对寻优过程实施动态监测，一旦发生早熟收敛则进行混沌搜索，即将优化变量通过载波的方式映射为混沌变量，并运用混沌优化机理继续在解空间中搜索[12]，以保持种群多样性，从而有效抑制早熟收敛。其中，早熟收敛判据为[13]

$$m < \frac{\sigma_{k+1}}{\sigma_k} < n \quad (12)$$

式中，$\sigma_{k+1}$ 和 $\sigma_k$ 分别为第 $k+1$ 和 $k$ 次迭代后的群体适应度方差值，本文中 $m$ 和 $n$ 分别取 0.99 和 1.01。

考虑到常用的 Logistic 映射遍历的不均匀性[12]，本文选取 Tent 映射为混沌映射。为了克服 Tent 映射迭代序列中存在小周期和不动点的缺陷，适时加入随机扰动，使 Tent 映射达到小周期或不动点时重新进入混沌状态。Tent 映射方程为

$$x_{k+1} = \begin{cases} 2x_k & 0 \leq x_k \leq 1/2 \\ 2(1-x_k) & 1/2 < x_k \leq 1 \end{cases} \quad (13)$$

当Tent映射达到小周期点（0.2, 0.4, 0.6, 0.8）或不动点（0, 0.25, 0.5, 0.75）时，首先加入随机扰动对当前变量 $x_{k+1}$ 更新，即

$$x_{k+1} = \frac{x_{k+1} + \mathrm{rand}(0,1)}{2} \quad (14)$$

然后再代入式（14）进行迭代。以初始值为 [0.534 6, 0.534 7]的二维向量为例，采用改进前后的Tent映射分别迭代5 000次，所得映射变量 $x_1$ 和 $x_2$ 分布如图1和图2所示。

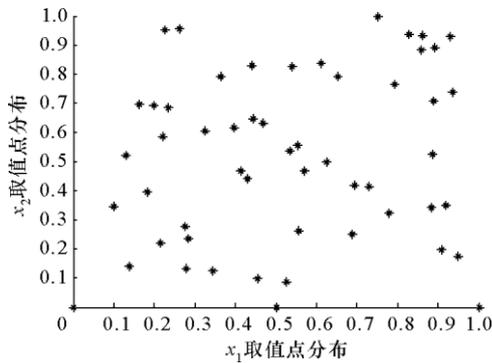

图1　Tent映射变量分布

Fig.1　Distribution of variables from the Tent map

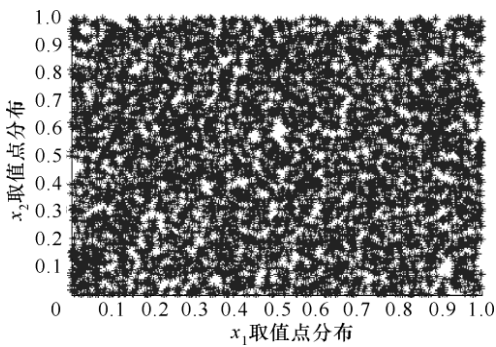

图2　改进的Tent映射变量分布

Fig.2　Distribution of variables from the improved Tent map

由图1可知，尽管迭代5 000次，但Tent映射所得解集数量远少于迭代次数。对所得数据进行分析可知，$x_1$ 和 $x_2$ 分别从第53和第51次迭代进入不动点后，所有后面取值均为0。

由图2可知，改进后的Tent映射变量分布密集且均匀，达到了预期效果。

### 3.2　适应度函数的确定

适应度函数是IBCC算法指导搜索方向的依据，因此寻优过程中构造一个合适的适应度函数非常重要。本文采用的适应度函数为训练集上的5-折交叉验证（Cross Validation，CV）分类正确率，即

$$F(\boldsymbol{u}) = F(\lambda, \sigma) = A_{5-\mathrm{cv}} \quad (15)$$

式中，$\boldsymbol{u}=[\lambda, \sigma]$ 为待优化的模型参数向量，在参数优化过程中用每个细菌个体的位置表示。

### 3.3　建模过程

基于IBCC的LLM模型参数优化的流程如图3所示。

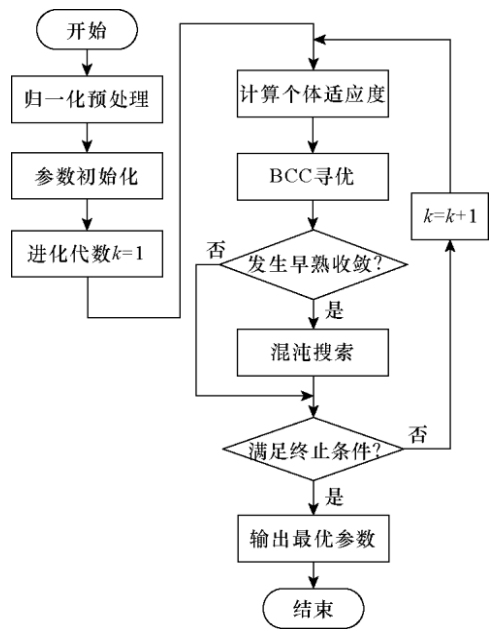

图3　参数优化流程图

Fig.3　Flowchart of parameter optimization

具体步骤如下：

（1）将训练样本和测试样本采用z-score规范化方法进行归一化预处理，即

$$l' = \frac{l - \overline{L}}{\sigma_L} \quad (16)$$

式中，$\overline{L}$、$\sigma_L$ 分别为样本集中任意特征量 $L$ 的均值和标准差，$l'$ 为 $L$ 中的数值 $l$ 被归一化后的值。

（2）初始化IBCC算法参数：种群规模为20，最大进化代数为200，初始精度为2，终止精度为0.000 01，精度更新参数为1.25，搜寻维数为2维，初始位置分别随机分布在区间(0, 500)和(0, 1 000)内，$S_{\min}$ 和 $S_{\max}$ 分别取1和取值区域内两点间最大距离，混沌搜索最大步数为200。

（3）根据式（15）评价各细菌个体的适应度。

（4）按照 BCC 算法的寻优机制，不断更新细菌的位置，同时根据式（10）和式（12）对群体适应度方差实施动态监测。

（5）一旦监测到发生早熟收敛，则保存当前最优解 $u^*$，并根据式（13）和式（14）进行混沌搜索。若在解空间内搜索到比 $u^*$ 更优的解 $u'$（即 $F(u')>F(u^*)$），则更新最优解 $u^*=u'$，退出混沌搜索，继续以 BCC 寻优。

（6）终止条件判断：判断进化代数是否达到最大进化代数或者适应度值大于 99.50%，若满足，则结束寻优，否则进化代数加 1，转至步骤（3）。

（7）得到优化的模型参数 $u^*=[\lambda^*,\sigma^*]$，由式（5）和式（9）得到暂态稳定评估的优化模型。

## 4 原始特征选择

TSA 的原始特征可以从以下角度进行分类：从是否随系统规模变化上分，原始特征可分为单机特征和系统特征[2]；从时间上分，原始特征可分为静态特征和动态特征[4]；从空间上分，原始特征可分为电网参数特征和发电机参数特征[4]。近年来，PMU 的引入使获取同步的故障后实时信息成为现实，从而为 TSA 提供了新的输入特征。

在原始特征集的构建过程中，本文依据了以下原则：①主流性原则：通过对电力系统暂态稳定物理过程的本质特性做深入分析，求取与稳定性强相关的输入特征量；②实时原则：基于引入 PMU 后的优势，求取来自扰动发生之后的信息，可以动态实时地表征故障发生后系统的运行状态。③系统性原则：选用系统特征，而不是单机特征，以保证输入变量个数不随系统规模的增大成比例增长，从而适合大系统的稳定分析。

本文按照上述原则，在综合现有的研究文献的基础上，通过大量仿真分析，构建了一组由 33 个系统特征所组成的原始特征集，见表 1 所示。表中 $t_0$ 为故障初始时刻，$t_{cl}$ 为故障切除时刻，$t_{cl+3c}$ 为故障切除后第 3 周波，$t_{cl+6c}$ 为故障切除后第 6 周波，$t_{cl+9c}$ 为故障切除后第 9 周波。

表 1 中，Tz1 反映了系统的整体静态稳定水平；Tz2~Tz4 为由故障瞬间抽取的特征量，其中 Tz2 反映了受扰最严重的发电机的失稳趋势，Tz3 反映了受扰最严重发电机的静态运行点，Tz4 反映了系统中各发电机故障瞬间供求关系失衡的平均水平；Tz5~Tz12 为由故障切除时刻抽取的特征量，其中

表 1　数据集的输入特征量
Tab.1　Input features of data set

| 编号 | 输入特征量 |
|---|---|
| Tz1 | 系统中各发电机机械功率的平均值 |
| Tz2 | $t_0$ 时刻所有发电机初始加速度的最大值 |
| Tz3 | $t_0$ 时刻具有最大加速度发电机的初始角度 |
| Tz4 | $t_0$ 时刻所有发电机初始加速功率的均值 |
| Tz5 | $t_{cl}$ 时刻系统冲击的大小 |
| Tz6 | $t_{cl}$ 时刻与惯性中心相差最大的发电机转子角度 |
| Tz7 | $t_{cl}$ 时刻具有最大转角发电机的动能 |
| Tz8 | $t_{cl}$ 时刻具有最大动能发电机的转子角度 |
| Tz9 | $t_{cl}$ 时刻所有发电机转子动能的最大值 |
| Tz10 | $t_{cl}$ 时刻所有发电机转子动能的平均值 |
| Tz11 | $t_{cl}$ 时刻发电机转子最大相对摇摆角 |
| Tz12 | $t_{cl}$ 时刻与惯性中心相差最大的发电机角速度 |
| Tz13 | $t_{cl+3c}$ 时刻系统冲击的大小 |
| Tz14 | $t_{cl+3c}$ 时刻所有发电机转子动能的最大值 |
| Tz15 | $t_{cl+3c}$ 时刻所有发电机转子动能的平均值 |
| Tz16 | $t_{cl+3c}$ 时刻与惯性中心相差最大的发电机转子角度 |
| Tz17 | $t_{cl+3c}$ 时刻发电机转子最大相对摇摆角 |
| Tz18 | $t_{cl+3c}$ 时刻具有最大转角发电机的动能 |
| Tz19 | $t_{cl+3c}$ 时刻与惯性中心相差最大的发电机角速度 |
| Tz20 | $t_{cl+6c}$ 时刻系统冲击的大小 |
| Tz21 | $t_{cl+6c}$ 时刻所有发电机转子动能的最大值 |
| Tz22 | $t_{cl+6c}$ 时刻所有发电机转子动能的平均值 |
| Tz23 | $t_{cl+6c}$ 时刻具有最大转角发电机的动能 |
| Tz24 | $t_{cl+6c}$ 时刻与惯性中心相差最大的发电机转子角度 |
| Tz25 | $t_{cl+6c}$ 时刻发电机转子最大相对摇摆角 |
| Tz26 | $t_{cl+6c}$ 时刻与惯性中心相差最大的发电机角速度 |
| Tz27 | $t_{cl+9c}$ 时刻系统冲击的大小 |
| Tz28 | $t_{cl+9c}$ 时刻具有最大转角发电机的动能 |
| Tz29 | $t_{cl+9c}$ 时刻所有发电机转子动能的最大值 |
| Tz30 | $t_{cl+9c}$ 时刻所有发电机转子动能的平均值 |
| Tz31 | $t_{cl+9c}$ 时刻与惯性中心相差最大的发电机转子角度 |
| Tz32 | $t_{cl+9c}$ 时刻发电机转子最大相对摇摆角 |
| Tz33 | $t_{cl+9c}$ 时刻与惯性中心相差最大的发电机角速度 |

Tz5 反映了故障切除时刻对系统的破坏情况，Tz7 反映了故障切除时转角最领先发电机的失稳趋势，Tz8 反映了具有最大动能发电机在故障切除后的减速能力；Tz13～Tz33 为考虑 PMU 可以提供故障后实测信息的优势而提出的特征量，它们反映了故障切除后的过程中系统的稳定特性。由此可知，上述所选特征量比较地全面表征了系统在故障前后的受扰过程中不同侧面、不同阶段的稳定特性、而彼此间又相互补充，故而选其构成所提方法的原始特征集。

## 5　算例与结果分析

### 5.1　算例介绍

新英格兰 10 机 39 节点系统[2,4,5,8]共由 10 台发电机、39 条母线和 46 条线路所组成，代表美国新英格兰州的一个 345 kV 电力网络，其中 39 号母线所连的发电机为外网等值机。

### 5.2　样本集的构造

同步发电机采用经典模型，负荷模型为恒阻抗模型；故障类型为三相短路，故障清除时间为 0.1s，故障清除后系统拓扑结构不变；系统在从 85%以 10%递增到 115%共计 4 个负荷水平下，每个负荷水平下随机设置 5 种发电机出力，选择 22 个不同的故障位置；失稳判据为仿真结束时，任意两台发电机的最大相对功角差是否大于 360°[4,5]；仿真软件为 PSD-BPA，共生成 440 个样本，随机选取其中 330 个构成训练集，其余为测试集。

### 5.3　结果与讨论

#### 5.3.1　参数优化结果比较

为验证 IBCC 的参数优化性能，本文将其与遗传算法（Genetic Algorithm，GA）、粒子群优化算法（Particle Swarm Optimization，PSO）和 BCC 算法等优化算法进行了对比测试。为便于比较，三种比较算法和 IBCC 中共有参数（如种群规模、最大迭代次数、搜索区间等）的取值均与 IBCC 相同，其他参数设置如下：GA 中交叉概率为 0.85，变异概率为 0.001；PSO 中学习因子 $c_1=c_2=2$，惯性权重从 0.9 线性下降到 0.4。考虑智能优化算法的随机性，各优化算法均独立运行 100 次，所得结果见表 2。表 2 中的搜索时间（在 Intel Pentium Dual CPU E2200 @2.20 GHz，2.19 GHz，1.0 GB 内存的 PC 机上）为运行 100 次的平均搜索时间。

表 2　不同优化算法的训练结果

Tab.2　Training results of different optimization algorithms

| 优化算法 | 搜索时间/s | 模型参数 $\lambda$ | 模型参数 $\sigma$ | 训练正确率(%) | 百次搜索成功率(%) |
|---|---|---|---|---|---|
| GA | 418.45 | 0.52 | 5.19 | 95.45 | 74 |
| PSO | 176.31 | 0.14 | 6.31 | 94.85 | 49 |
| BCC | 87.59 | 0.48 | 3.57 | 96.06 | 61 |
| IBCC | 96.74 | 0.25 | 5.03 | 99.39 | 93 |

由表 2 可知，四种优化算法均能有效优化选取 LLM 模型的参数。由于算法自身寻优机制的原因，相对于 GA 及 PSO，BCC 和 IBCC 的优化结果更好些，表现在训练正确率更高、搜索时间更短、搜索成功率更高。同时，IBCC 算法由于采用了感知范围自适应调整机制，并需要对寻优过程实施动态监测，在发生早熟收敛时进行混沌搜索，所以其搜索到的最优结果最好，性能最稳定（百次搜索成功率最高）。

在参数优化过程中，四种优化算法的最佳个体对应的训练正确率进化曲线如图 4 所示。

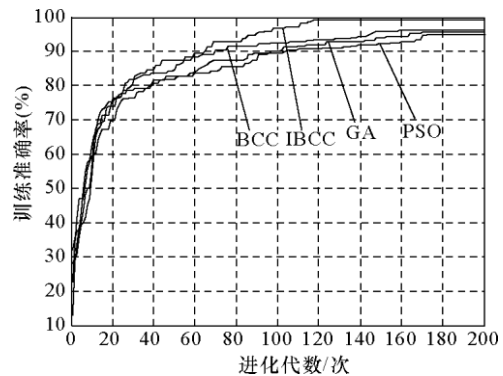

图 4　几种算法的训练正确率进化曲线

Fig.4　Training accuracy curves of different algorithms

由图 4 可以看出，在 LLM 模型参数优化问题上，四种优化算法均取得了明显效果。其中，IBCC 的收敛速度最快，在第 118 代达到最优；同时 IBCC 在第 70 代附近曾出现短暂的停顿，但很快又继续爬升，说明 IBCC 凭借其强大的全局搜索能力跳出这一局部最优解，从而验证了 IBCC 算法的有效性。

#### 5.3.2　测试结果比较

基于所得最优模型对测试样本进行稳定评估，所得测试结果如表 3 所示。同时，可得原始特征集中特征权重如图 5 所示。

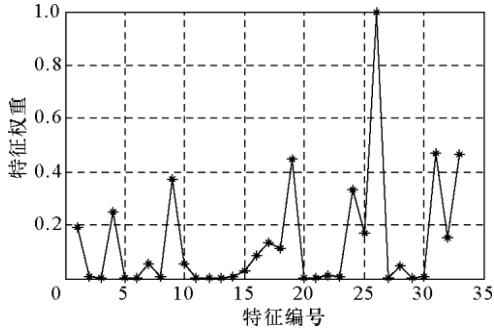

图 5　特征权重

Fig.5　Feature weights

由表 3 可知，所提方法的测试正确率最高，达到 99.09%，从而验证了 IBCC 算法能有效优化 LLM 模型的参数，提高评估的分类正确率。

表 3　新英格兰 10 机 39 节点系统测试结果

Tab.3　Test results in New England test system

| 测试系统 | 评估方法 | 测试正确率(%) |
|---|---|---|
| 新英格兰 10 机 39 节点系统 | GA-LLM | 95.45 |
|  | PSO-LLM | 94.55 |
|  | BCC-LLM | 96.36 |
|  | IBCC-LLM | 99.09 |

由图 5 可知，原始特征集中各个特征的权重差别较大，这表明不同输入特征对稳定结果贡献的"价值"高低不同。因此，所提方法在进行暂态稳定评估的同时，也能进行特征选择、并给出各个特征分类"价值"的定量评价，为 TSA 的输入特征选择和输入空间降维提供了参考。

**5.3.3　算法的通用性分析**

为了验证所提方法的通用性，本文设计了两组实验，研究当分别缺少若干（如 5 个）权重较小和较大的特征量时，对计算结果的影响情况，并用 SVM 进行了对比，结果见表 4。其中 SVM 的核函数选用径向基核函数，算法参数通过网格搜索结合 5-折交叉验证优化选取[5]。

表 4　缺少部分特征量时的评估结果

Tab.4　Test results of lacking some features

| 实验方案 | 所缺特征量 | 评估模型 | 测试正确率(%) |
|---|---|---|---|
| 实验 1 | Tz21, Tz5, Tz14, Tz15, Tz10 | 本文方法 | 99.09 |
|  |  | SVM | 98.18 |
| 实验 2 | Tz26, Tz24, Tz19, Tz1, Tz25 | 本文方法 | 98.18 |
|  |  | SVM | 95.45 |

由表 4 可知，缺少若干特征量会对评估结果造成不同程度的影响。当所缺特征量的权重较小时，计算结果所受影响较小；而当所缺特征量的权重较大时，计算结果所受影响较大。这表明，特征量选取的准确性对评估正确率有重要影响。同时，在以上两个实验中，本文方法的测试正确率均高于 SVM，特别在实验 2 中本文方法的测试正确率比 SVM 高出 2.73%。这说明 IBBC-LLM 方法通过特征加权策略能自动调整特征的权重，对特征量变化具有良好的适应性。

**5.4　其他暂态稳定评估模型的评估结果**

**5.4.1　不加无关特征时的评估结果**

为验证所提方法的有效性，本文选用多层感知器（Multilayer Perception，MLP）[2]、SVM[4-5]、DT[3]、PNN[6]和 ELM[7]等评估模型进行了对比测试，测试结果见表 5。其中，各评估模型参数设置如下：MLP 为单隐层网络，隐层神经元个数为 25，训练算法为反向传播算法，学习率 0.8，动量因子为 0.7；SVM 的参数设置同 5.3.3 节；DT 采用 C 4.5 算法，置信因子设定为 0.25；PNN 的径向基函数传播系数为 0.1；ELM 的隐层节点个数为 50。

表 5　其他模型不加无关特征时的测试结果

Tab.5　Test results of other models without irrelevant features

| 测试系统 | 评估模型 | 测试正确率(%) |
|---|---|---|
| 新英格兰 10 机 39 节点系统 | MLP | 95.45 |
|  | SVM | 96.36 |
|  | DT | 95.45 |
|  | PNN | 97.27 |
|  | ELM | 98.18 |

由表 5 可知，不加无关特征时，MLP、SVM 和 DT 等评估模型均能有效评估电力系统的暂态稳定性，测试正确率均在 95% 以上，其中 ELM 算法的评估结果与所提方法相近，达到了 98.18%。

**5.4.2　加入无关特征时的评估结果**

为了检验所提方法在含有无关特征时的性能，在原始特征集中依次加入 0~200 个无关特征，所得测试正确率见表 6。其中，所添加的各无关特征的变量取值是由服从高斯分布（均值为 0、方差为 1）的随机变量所生成的。

**表 6 所提方法含有无关特征时的测试结果**

Tab.6 Test results of our method containing irrelevant features

| 测试系统 | 无关特征数量 | | | | |
|---|---|---|---|---|---|
| | 0 | 50 | 100 | 150 | 200 |
| 新英格兰 10 机 39 节点系统 | 99.09 | 99.09 | 99.09 | 99.09 | 99.09 |

从表 6 可知，随着无关特征数量的增加，所提算法的测试正确率保持不变，这表明所提算法对逐渐增多的无关特征不敏感。

当含有 200 个无关特征时，所提方法与 MLP、SVM 和 DT 等评估模型的测试结果见表 7。

**表 7 不同模型含有 200 个无关特征时的测试结果**

Tab.7 Test results of different models containing 200 irrelevant features

| 测试系统 | 评估模型 | 测试正确率（%） |
|---|---|---|
| 新英格兰 10 机 39 节点系统 | MLP | 75.45 |
| | SVM | 85.45 |
| | DT | 76.36 |
| | PNN | 83.64 |
| | ELM | 90.91 |
| | IBCC-LLM | 99.09 |

从表 7 可知，相对于不加无关输入特征，当输入特征中含有较多无关输入特征时，MLP、SVM、DT、PNN 和 ELM 等评估模型的测试正确率均有下降，这表明这些评估模型均不同程度地受到"维数灾"影响，其中 MLP 和 DT 二者受到的影响最大、测试正确率下降最为明显，而所提方法的测试正确率仍高达 99.09%，明显优于其他评估模型。分析其原因，这是因为 LLM 算法基于大间隔理论，实现了两个层次的正则化——隐式的留一法正则化和显式的 $\ell$ 正则化[9]，从而使所提算法在加入大量无关特征时仍保持优越的性能。

## 6 结论

为了提高电力系统暂态稳定评估的分类正确率，提出一种基于局部学习机和改进细菌群体趋药性算法的暂态稳定评估方法，在新英格兰 10 机 39 节点系统上进行了仿真研究，结论如下：

（1）改进 BCC 算法能有效优化 LLM 模型的参数，提高评估的分类正确率，而且比 GA、PSO 和标准 BCC 等优化算法具有更稳定的性能、更强的全局寻优能力和更快的收敛速度。

（2）所提方法比 MLP、SVM、DT、PNN 和 ELM 等其他暂态稳定评估模型具有更高的分类正确率，而且算法性能对无关特征数量的增加不敏感。

（3）所提方法在进行暂态稳定评估的同时，也能进行特征选择、并给出各个输入特征的分类"价值"的定量评价，从而为 TSA 的输入特征选择和输入空间降维提供了参考。

作者简介

顾雪平　男，1964 年生，博士，教授，博士生导师，主要研究方向为电力系统安全防御和系统恢复、电力系统安全稳定评估与控制、智能技术在电力系统中的应用。

李　扬　男，1980 年生，博士研究生，主要研究方向为电力系统安全稳定评估与控制、智能技术在电力系统中的应用。